\documentclass[a4paper,conference]{IEEEtran}

\usepackage{ifthen}
\usepackage{amsfonts, amssymb, amsthm, amsopn}
\usepackage{amsmath}
\usepackage{mathrsfs}
\def\atauxout{\csname @auxout\endcsname}%
\def\labelii#1{\immediate\write\atauxout{%
    \noexpand\newlabel{#1}{{\theenumii}{\thepage}}}}

\newcommand{\bfx}{\vect{x}}
\newcommand{\bfX}{\vect{X}}

\newcommand{\vect}[1]{\mathbf{#1}} 
 
\newcommand{\Reals}{\mathbb R}      
\newcommand{\Integers}{\mathbb Z}   
                          
\newcommand{\const}[1]{\textnormal{\usefont{U}{eur}{m}{n}\selectfont #1}}

\newcommand{\amp}{\const{A}}

\newcommand{\TT}{\const{T}}

\newcommand{\MM}{\const{M}}  

\newcommand{\Prvcond}[2]{\Pr[#1 \kern0.1em|\kern0.1em #2]} 
\newcommand{\bigPrvcond}[2]{\Pr\bigl[#1 \kern-0.1em \bigm| \kern-0.1em#2\bigr]}
\newcommand{\BigPrvcond}[2]{\Pr\Bigl[#1 \kern-0.1em \Bigm| \kern-0.1em#2\Bigr]}
\newcommand{\biggPrvcond}[2]{\Pr\biggl[#1 \kern-0.1em \biggm| \kern-0.1em#2\biggr]}

\newcommand{\Prscond}[2]{\Pr(#1 \kern0.1em|\kern0.1em #2)} 
\newcommand{\bigPrscond}[2]{\Pr\bigl(#1 \kern-0.1em \bigm| \kern-0.1em#2\bigr)}
\newcommand{\BigPrscond}[2]{\Pr\Bigl(#1 \kern-0.1em \Bigm| \kern-0.1em#2\Bigr)}

\newcommand{\biggPrscond}[2]{\Pr\biggl(#1 \kern-0.1em \biggm| \kern-0.1em#2\biggr)}

\newcommand{\Econd}[3][]{\textnormal{\textsf{E}}_{#1}[#2
    \kern0.1em|\kern0.1em #3]}
\newcommand{\bigEcond}[3][]{\textnormal{\textsf{E}}_{#1}\!\bigl[#2
  \kern-0.1em \bigm| \kern-0.1em #3\bigr]}
\newcommand{\BigEcond}[3][]{\textnormal{\textsf{E}}_{#1}\!\Bigl[#2
  \kern-0.1em\Bigm|\kern-0.1em #3\Bigr]}
\newcommand{\biggEcond}[3][]{\textnormal{\textsf{E}}_{#1}\!\biggl[#2
  \kern-0.1em \biggm| \kern-0.1em #3\biggr]}
\newcommand{\BiggEcond}[3][]{\textnormal{\textsf{E}}_{#1}\!\Biggl[#2
  \kern-0.1em \Biggm| \kern-0.1em #3\Biggr]}
\newcommand{\BigCovcond}[3]{\textnormal{\textsf{Cov}}\!\Bigl[
  {#1},{#2} \kern-0.1em \Bigm| \kern-0.1em {#3} \Bigr]}

\newcommand{\bigCovcond}[3]{\textnormal{\textsf{Cov}}\!\bigl[
  {#1},{#2} \kern-0.1em \bigm| \kern-0.1em {#3} \bigr]}

\newcommand{\E}[2][]{\textnormal{\textsf{E}}_{#1}\!\left[#2\right]} 

\renewcommand{\d}{\,\textnormal{d}}

\newcommand{\I}[1]{\operatorname{I}\{#1\}}

\newcommand{\eps}{\epsilon} 

\newtheorem{theorem}{Theorem} 

\begin{document}
\title{It Takes Half the Energy of a Photon to Send One Bit Reliably on the
  Poisson Channel with Feedback}


%


\author{%
\IEEEauthorblockN{Shahab Asoodeh, Amos Lapidoth,  and Ligong Wang}
\IEEEauthorblockA{ETH Zurich\\
             ISI (D-ITET), Sternwartstr. 7\\
              CH-8092 Z\"urich, Switzerland\\
              Email: lapidoth@isi.ee.ethz.ch}
}



\maketitle

\begin{abstract}
  We consider the transmission of a single bit over the
  continuous-time Poisson channel with noiseless feedback.  We show
  that to send the bit reliably requires, on the average, half the
  energy of a photon. In the absence of peak-power constraints this
  holds irrespective of the intensity of the dark current. We also
  solve for the energy required to send $\log_{2} \MM$ bits.
\end{abstract}

\section{Introduction}

The continuous-time Poisson channel models optical communication using
direct detection. The input to the channel $x(\cdot)$ is nonnegative
\begin{equation}
  x(t) \geq 0, \quad t \in \Reals,
\end{equation}
and conditional on the input, the output $Y(\cdot)$ is a conditional
Poisson process (also known as a doubly-stochastic Poisson process) of
intensity $x(t) + \lambda_{0}$, where $\lambda_{0}$ is a nonnegative
constant called \emph{dark current}. Thus, conditional on the input,
the output $Y(\cdot)$ is a nonhomogeneous Poisson process and thus of
independent increments with
\begin{equation}
  \bigPrvcond{Y(t+\tau) - Y(t) = \nu}{\bfX = \bfx} = e^{-\Lambda}
  \frac{\Lambda^{\nu}}{\nu !}, \quad \nu \in \Integers,
\end{equation}
where
\begin{equation}
  \Lambda = \int_{t}^{t + \tau} \bigl( x(\sigma) + \lambda_{0} \bigr)
  \d{\sigma}. 
\end{equation}

To send a bit $D$ taking on the values $0$ and $1$ equiprobably over
this channel \emph{without feedback} we use two input waveforms
$x_{0}(\cdot)$ and $x_{1}(\cdot)$, and we send $x_{0}(\cdot)$ if $D=0$
and $x_{1}(\cdot)$ if $D=1$.

We refer to
\begin{equation*}
  \E{\int_{t}^{t + \tau} X(\sigma) \d{\sigma}}
\end{equation*}
as the transmitted energy in the time interval $[t, t+\tau]$, although
this is somewhat imprecise: this quantity is the expected number of
transmitted photons in the interval, and one should technically
multiply it by the energy in each photon (which depends on the light
frequency) to obtain the transmitted energy in the interval.

We sometimes impose a \emph{peak-power constraint} on the input, in
which case we require that, with probability one,
\begin{equation}
  x(t) \leq \amp.
\end{equation}
We then refer to $\amp$ as the maximal allowed power (although,
technically speaking, this needs to be normalized by the energy of
each photon to have the sense of power.)

In the presence of feedback, the channel description is a bit more
technical \cite{Lap93}. We require that conditional on $D=0$, the
channel output $Y(t)$ admit the $\mathcal{F}_{t}$ intensity
$X_{0}(t) + \lambda_{0}$ \cite[Chapter~II, Section~3, Definition
D7]{Bremaud}. That is, conditional on $D=0$, $Y(t)$ is a point process
adapted to some history ${\mathcal F}_{t}$; $X_{0}(t)$ is a
nonnegative ${\mathcal F}_{t}$-progressive process such that for all
$t \geq 0$
  \begin{equation*}
    \int_{0}^{t} X_{0}(s) \d{s} < \infty;
  \end{equation*}
  and for all nonnegative ${\mathcal F}_{t}$-predictable processes $C(t)$
  \begin{equation}
    \label{eq:cryptic}
    \E{ \int_{0}^{\infty} C(t) \d Y(t)} = \E{ \int_{0}^{\infty} C(t)
      \bigl(X_{0}(t) + \lambda_{0}\bigr) \d{t}}.
  \end{equation}
  The conditional expected energy transmitted when $D=0$ over the time
  interval $[0,\TT]$ is
  \begin{equation}
    {\mathcal E}_{0} = \E{ \int_{0}^{\TT} X_{0}(t) \d{t}}.
  \end{equation}
  Similarly, when $D=1$ the transmitted energy is
\begin{equation}
    {\mathcal E}_{1} = \E{ \int_{0}^{\TT} X_{1}(t) \d{t}}.
  \end{equation}
The average transmitted energy is thus
\begin{equation}
  \frac{1}{2} \bigl( {\mathcal E}_{0} + {\mathcal E}_{1} \bigr).
\end{equation}
A decoder is a mapping from the $\sigma$-algebra generated by $\{Y(t),
\; 0 \leq t \leq \TT\}$ to the set $\{0,1\}$.

We say that a bit can be transmitted reliably over our channel with
average transmitted energy ${\mathcal E}$, if for any $\eps > 0$ we
can find some transmission interval $\TT$ and a coding/decoding rule
of expected transmission energy ${\mathcal E}$ and probability of
error smaller than $\eps$. We denote by ${\mathcal E}_{\text{min}}$ the
  least energy required to transmit a bit reliably over our channel.

\section{Main Result}

\begin{theorem}
  The minimum energy required to send a single bit over the Poisson
  channel with dark current $\lambda_{0}$, feedback, and no peak-power
  constraint is
  \begin{equation}
    {\mathcal E}_{\textnormal{min}} = \frac{1}{2},
  \end{equation}
  irrespective of the dark current. If the dark current is zero, then
  this is achievable even under a peak-power constraint whenever $\amp
  > 0$.
\end{theorem}

\section{Coding Scheme}

The achievability when $\lambda_{0} = 0$ is straightforward. To send
$D=0$ we transmit the all-zero input; to send $D=1$ we transmit $\amp$
until, through the feedback link, we learn that a count was
registered; thereafter we send zero. The decoder guesses ``$D=0$'' if
no counts were registered in the interval $[0,\TT]$, and guesses
``$D=1$'' otherwise. Since $\lambda_{0} = 0$, the probability of error
given $D=0$ is zero. Also, the transmitted energy when $D=0$ is zero,
so ${\mathcal E}_{0} = 0$. Conditional on $D=1$ the time of the first
count is exponential with mean $1/\amp$. Consequently, ${\mathcal
  E}_{1} = 1$. Conditional on $D=1$, the probability of error is the
probability that no counts are registered in the interval
$[0,\TT]$. The probability of this event is the probability that the
first count occurs after time $\TT$, i.e., the probability that a
mean-$1/\amp$ exponential exceeds $\TT$. It thus tends to zero as $\TT
\to \infty$.

If $\lambda_{0} > 0$ and there is no peak-power constraint, we choose
$\amp \gg 1$ and $\Delta \ll 1$ and use the above scheme with $\TT =
\Delta$. We make sure that $\Delta$ is small enough for the
probability of a spurious count in the interval $[0,\Delta]$ to be
very small (the probability of a spurious count in this interval is
$1- e^{-\Delta \lambda_{0}}$), and we choose $\amp$ large enough so
that the probability that a mean-$1/\amp$ exponential exceeds $\Delta$
is also very small.

\section{Converse}

To prove that ${\mathcal E}_{\text{min}}$ cannot be smaller than
$1/2$, it suffices to consider the case where $\lambda_{0} = 0$. We
thus assume $\lambda_{0} = 0$. In this case there is no loss in
optimality in assuming that to send $D=0$ we transmit the all-zero
input. Indeed, given any general scheme consider the guess the decoder
produces when faced with no counts. Call that FALSE. Let TRUE be its
complement.  Consider now a scheme with the same encoding rule for
TRUE, with the same decoding rule, but where we send the all-zero
input to convey FALSE.  The new scheme uses less (or same) energy; has
the same $p(\text{error}|\text{TRUE})$; and has
$p(\text{error}|\text{FALSE}) = 0$. Since the name we give to the
hypotheses is immaterial, we can assume that FALSE corresponds to
$D=0$.

Next we argue that there is no loss in optimality in restricting
ourselves to a detector that bases its decision on the presence of
counts in the interval $[0,\TT]$. Since this enlarges the set of
outcomes yielding the guess ``D=1'', this cannot increase
$p(\text{error}|D=1)$.  To send $D=0$ we send the all-zero waveform,
which results in no counts (there is no dark current), so this does not
change $p(\text{error}|D=0)$.

Finally, we argue that there is no loss in optimality in stopping
transmission once a count has been registered. Indeed, this reduces
the transmitted energy and does not change the performance of the
above detector. 

We next analyze the probability of error of such schemes. Conditional
on $D=0$, the probability of error is zero, because there is no dark
current so sending zero input guarantees zero counts. As to the
conditional probability of error given $D=1$, let $T_{1}$ denote the
random time at which the first count is registered.  Substituting the
stochastic process
\begin{equation}
  C(\omega, t) = \I{ t \leq T_{1}(\omega) \wedge \TT}, \quad (\omega,t)
  \in \Omega \times [0,\infty).
\end{equation}
in \eqref{eq:cryptic} yields
\begin{align*}
  p(\text{correct}|D=1) & = \Prvcond{T_{1} \leq \TT}{D=1} \\
  & = \bigEcond{Y(T_{1} \wedge \TT)}{D=1} \\
  & = \biggEcond{ \int_{0}^{\infty} C(t) \d Y(t)}{D=1} \\
  & = \E{ \int_{0}^{\infty} C(t) X_{1}(t) \d{t}} \\
  & = \E{ \int_{0}^{T_{1} \wedge \TT} X_{1}(t) \d{t}} \\
  & = {\mathcal E}_{1}.
\end{align*}
For the probability of error to tend to zero, the expected energy
transmitted to convey $D=1$ must thus approach $1$.

\section{Sending $\log_{2} \MM$ Bits}
More generally, to send $\log_{2} \MM$ bits requires
\begin{equation}
  \label{eq:sendK}
  \frac{\MM - 1}{\MM} 
\end{equation}
of the energy of a photon. When $\MM=2$ we recover the required energy
to send one bit. To prove the converse---that one cannot accomplish
the task with less energy---requires a simple genie-aided
argument. Once again we can assume no dark current, and we can
show that there is no loss in optimality in conveying the zero message
using the all-zero input and by limiting ourselves to detectors that
guess that the transmitted message was the zero message if, and only
if, no counts were registered. We then consider a genie that, if a
count is registered, tells the detector which message was sent. With
the aid of the genie the detectors errs if, and only if, the
transmitted message was not the zero message and no counts were
registered. Thus, by our previous analysis, the required energy of
each of the nonzero messages must be one. Averaging over the
equally-likely messages demonstrates that $(\MM-1)/\MM$ of the
energy of a photon is necessary for reliable transmission.

The direct part in the absence of dark current is based on a simple
scheme where the transmission interval $[0,\TT)$ is divided into
$\MM-1$ intervals. We associate with the nonzero message~$m \in
\{1,\ldots, \MM-1\}$ the interval
\begin{equation*}
  \bigl[ (m-1) \TT/(\MM - 1), m \TT/(\MM-1) \bigr).
\end{equation*}
To send the nonzero message~$m$ we transmit with a very high power
starting at time~$(m-1) \TT/(\MM - 1)$ until a count is registered or
until the end of the interval at time $m \TT/(\MM-1)$. To send the
zero message we send the all-zero signal.  The detector operates as
follows. If no counts are registered, it guesses that the zeroth
message was sent. If a count is registered, it declares that the
transmitted message was the one that corresponds to the interval in
which the count was registered.

In the presence of dark current we use the same scheme except that
we choose as our transmission time a very short interval $[0,\Delta]$
in which the probability of a spurious count is negligible.

\section{Discussion}

In the absence of a peak power constraint, the capacity of the Poisson
channel (with or without feedback) is infinite \cite{kabanov78_1},
\cite{davis80_1}. Thus, in sending a very large number of bits,
reliable communication can be had with an arbitrarily small expected
energy per bit. The situation changes dramatically when sending a
\emph{single} bit. Even in the presence of feedback, the required
energy is finite; it is, in fact, $1/2$.

This should be contrasted with the infinite-bandwidth Gaussian channel,
where a single bit can be sent reliably with the same amount of energy
that would be required per bit if one were sending a large number of
bits \cite{Turin}, \cite{Poly}.

However, if the allowed energy is that of one photon, then we can send
as many bits as we want with arbitrarily small probability of error.


\end{document}